 \documentclass[twocolumn]{aastex6}
 \usepackage{amsmath}

\fullcollaborationName{The Friends of AASTeX Collaboration}

\begin{document}


\title{Effect of spectral index distribution on estimating the AGN radio luminosity function}


\author{Zunli Yuan\altaffilmark{1,2}, Jiancheng Wang\altaffilmark{1,2}, Ming Zhou\altaffilmark{1,2} and Jirong Mao\altaffilmark{1,2}}
\affil{yuanzunli@ynao.ac.cn}







\altaffiltext{1}{Yunnan Observatories, Chinese Academy of Sciences,  Kunming 650011, China}
\altaffiltext{2}{Key Laboratory for the Structure and Evolution of Celestial Objects, Chinese Academy of Sciences,  Kunming 650011, China}

\begin{abstract}

In this paper, we scrutinize the effect of spectral index distribution on estimating the AGN (active galactic nucleus) radio luminosity function (RLF) by a Monte Carlo method. We find that the traditional bivariate RLF estimators can cause bias in varying degree. The bias is especially pronounced for the flat-spectrum radio sources whose spectral index distribution is more scattered. We believe that the bias is caused because the $K$-corrections complicate the truncation boundary on the $L-z$ plane of the sample, but the traditional bivariate RLF estimators have difficulty in dealing with this boundary condition properly. We suggest that the spectral index distribution should be incorporated into the RLF analysis process to obtain a robust estimation. This drives the need for a trivariate function of the form $\Phi(\alpha,z,L)$ which we show provides an accurate basis for measuring the RLF.

\end{abstract}


\keywords{galaxies: active --- galaxies: luminosity function, mass function --- radio continuum: galaxies.}



\section{Introduction}

The luminosity function (LF), which provides a census of the galaxy and active galactic nucleus (AGN) populations over cosmic time, has been an important and also common tool for understanding the evolution of galaxies and ANGs \citep[][]{2008ApJ...682..874K}. With the abundance of multi-wavelength observed data for AGNs, AGN LFs have been estimated for various wave bands, such as the optical LF \citep[OLF, e.g.,][]{2000MNRAS.317.1014B}, the X-ray LF \citep[XLF, e.g.,][]{2000A&A...353...25M}, the $\gamma$-ray LF \citep[GLF, e.g.,][]{2012ApJ...751..108A,2013MNRAS.431..997Z} and the radio LF \citep[RLF, e.g.,][]{2001MNRAS.322..536W,2016ApJ...820...65Y,1990MNRAS.247...19D}. In this work, we will focus on the AGN RLFs.

In an actual survey, only a very limited number of objects in the universe can be observed. How many sources entering the sample depends on the survey depth and selection function. Thus the estimation of LFs is inevitably based on a truncated sample of objects. Another difficulty is brought about by $K$-correction. It not only affects the accurate determination of intrinsic luminosity of individual sources, but also complicates the process of translating flux selection limits into luminosity selection limits, even for a single band selected survey \citep{2007ApJ...661..703S,2016arXiv160407493L}. The truncation boundary on the $L-z$ plane of a real sample is often a complicated region, but not a regular curve. Therefore, $K$-correction can affect the estimation of LFs by making it difficult to define the truncation boundary. For example, \citet{2004MNRAS.351..541I} found that a wide range of $K$-corrections being applied across different galaxy types can bias the shape of the global LF.

For the radio AGNs, their spectra are frequently characterized as a simple power-law, $S\propto\nu^{-\alpha}$, and the $K$-correction has a simple form of $K(z)=(1+z)^{1-\alpha}$. Then the $K$-corrections of radio sources can be represented by their spectral properties (include spectral curvature, spectral index $\alpha$ and its distribution). Several Authors have discussed the potential problem of spectral curvature and its effect on obtaining reliable K-corrections for distant flat-spectrum radio sources \citep{1985MNRAS.217..601P,1996Natur.384..439S,2005A&A...434..133W} and also for steep-spectrum sources \citep{2011MNRAS.416.1900R}. Particularly, \citet{2000MNRAS.319..121J} highlighted the effect of spectral curvature that removes the evidence for the rapid decline in number density at high redshift suggested by \citet{1996Natur.384..439S}, and suggested that curvature would need to be incorporated in a full analysis of the RLF. Nevertheless, a recent study of \citet{2012MNRAS.422.2274C} pointed out that the effect of curvature only becomes important at higher frequencies ($\nu > 5 GHz$) and it can be avoided by measuring the spectra at lower frequencies. In the same paper, \citet{2000MNRAS.319..121J} also highlighted the importance of a distribution in spectral index in the parametric modeling of RLF \citep[also see][]{2001MNRAS.327..907J}. In this work, we analyze the bias caused by traditional bivariate RLF estimators based on Monte Carlo simulations, and further prove the necessity of incorporating the spectral index distribution into the RLF calculation.

Throughout the paper, we adopt a Lambda Cold Dark Matter cosmology with the parameters $\Omega_{m}$ = 0.27,  $\Omega_{\Lambda}$ = 0.73, and $H_{0}$ = 71 km s$^{-1}$ Mpc$^{-1}$.

\section{Methods}
The procedure of our method is summarized as follows. Firstly, simulated samples of radio AGNs are generated by the Monte Carlo method at given input RLFs. Then the traditional bivariate RLF estimators and a trivariate estimator that considers the spectral index distribution are respectively used to estimate the RLFs based on the simulated samples. Finally, the estimated RLFs are compared with the input RLFs to quantify the effect of spectral index distribution on estimating the RLF.

\subsection{The trivariate RLF}

To begin with, we need to define a trivariate RLF in which the distribution of spectral index is incorporated. It is defined as the number of sources per comoving volume $V(z)$ with radio luminosities in the range $L,L+dL$, and with spectral indexes in the range $\alpha,\alpha+d\alpha$.

\begin{eqnarray}
\label{aaa}
\Phi(\alpha,z,L)=\frac{d^{3}N}{d\alpha dz dL}.
\end{eqnarray}

\begin{table*}
\tablewidth{0pt}
\renewcommand{\arraystretch}{1.0}
\caption{Parameters}
\begin{center}
\begin{tabular}{lccccccccccccc}
\hline\hline

\colhead{     }             &
\colhead{     }             & \colhead{$\log_{10}\phi_0$} & \colhead{$\log_{10}\phi_1$} &
\colhead{$\log_{10}L_*$}    & \colhead{p$_1$} &
\colhead{p$_2$}           & \colhead{m} &
\colhead{z$_{\rm 0}$}       & \colhead{z$_{\rm \sigma}$} &
\colhead{k$_1$}             & \colhead{k$_2$}    &
\colhead{$\mu$}             & \colhead{$\sigma$} \\

\hline
               & Input value  & -4.20   & -4.44&  25.86 & 1.76 & 0.53 & -0.73 & 0.91 & 0.32 & -0.11 & 0.80 & 0.75& 0.23\\

Steep-spectrum & 2D fitting   & -       & -4.47&  25.89 & 1.77 & 0.59 & -0.55 & 0.93 & 0.34 & -0.10 & 0.74 & -   & -   \\

               & 3D fitting   & -4.13   & -4.44&  25.81 & 1.74 & 0.53 & -0.73 & 0.92 & 0.32 & -0.11 & 0.79 & 0.75& 0.23\\
\hline
               & Input value  & -5.60   & -5.58&  26.40 & 1.76 & 0.53 & -0.73 & 0.91 & 0.32 & -0.11 & 0.80 & 0.00& 0.42\\

Flat-spectrum  & 2D fitting   &    -    & -6.05&  26.61 & 2.24 & 0.47 & -1.12 & 0.86 & 0.29 & -0.13 & 0.96 & -   & -   \\

               & 3D fitting   & -5.60   & -5.58&  26.41 & 1.76 & 0.54 & -0.75 & 0.91 & 0.32 & -0.11 & 0.79 & 0.00& 0.42\\
\hline

\end{tabular}
\end{center}
~~Units -- $\phi_0$ and $\phi_1$: [${\rm Mpc^{-3}}$],\,\, $L_*$: [${\rm W Hz^{-1}}$]. The parameter values of the input RLFs and the best-fitting parameters.
\label{tab:fit}
\end{table*}

If the spectral index is independent of redshift and luminosity, $\Phi(\alpha,z,L)$ can be written as
\begin{eqnarray}
\label{aaa1}
\frac{d^{2}N}{dL dV}\times\frac{dN}{d\alpha}\times\frac{dV}{dz}=\rho(z,L)\times\frac{dN}{d\alpha}\times\frac{dV}{dz}.
\end{eqnarray}
where $\rho(z,L)$ is the common defined RLF (or referred as bivariate RLF), and $dV/dz$ is the co-moving volume element. The function $dN/d\alpha$ is the intrinsic spectral index distribution. $\Phi(\alpha,z,L)$ is related to the probability distribution of $(\alpha,z,L)$ by

\begin{eqnarray}
\label{eqpc}
p(\alpha,z,L)=\frac{1}{N_{tot}}\Phi(\alpha,z,L)\frac{dV}{dz}.
\end{eqnarray}

where $N_{tot}$ is the total number of sources in unit solid angle in the universe, and is given by the integral of $\Phi$ over $\alpha$, $L$ and $V(z)$. The probability function $p$ can be used to generate random draws of ($\alpha,z,L$) by the Monte Carlo method, once we assume a parametric form for $\Phi(\alpha,z,L)$.

\begin{figure}[!htb]
\centering
\includegraphics[width=1.0\columnwidth]{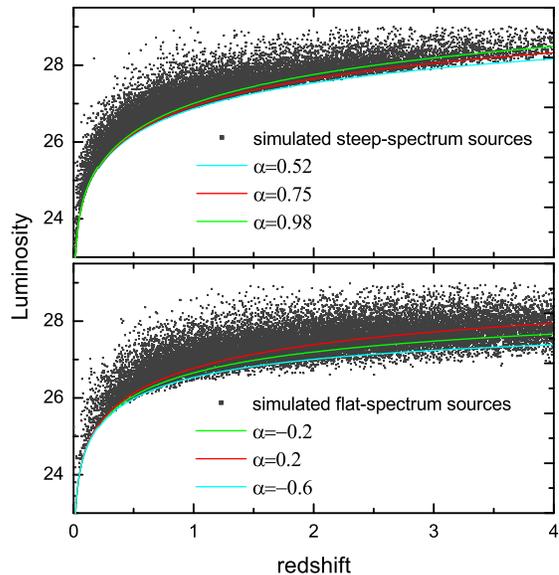}
\caption{Two sub-samples from our simulated steep- and flat-spectrum flux limited samples,respectively. Each of them contains about 20,000 sources, and the flux limit is 0.2 Jy. The colored solid lines show the flux limit curves $L=L_{lim}(\alpha,z)$ (also known as truncation boundary) with different $\alpha$.}
\label{simulated_samples}
\end{figure}

\subsection{The input RLF}

The input RLF we used is the model C of \citet{2016ApJ...820...65Y}.
\begin{eqnarray}
\label{eqn:rhoform}
\rho(z,L)=e_1(z)\rho(z=0,L/e_2(z)),
\end{eqnarray}
where $e_1(z)$ and $e_2(z)$ are the functions describing the density and the luminosity evolution respectively.
\begin{eqnarray}\label{eqn:LFPLEnoev}
e_1(z) =
\begin{cases}
  \displaystyle z^m\exp\left[-\frac{1}{2}\left(\frac{z-z_0}{z_{\sigma}}\right)^2\right],  &  0<z \leqslant z_0 \\
  z^m, & z>z_0
\end{cases}
\end{eqnarray}
The local luminosity function $\rho(z=0,L/e_2(z=0))$ has a double power law form described as:
\begin{eqnarray}
\label{local}
\begin{aligned}
\phi_1\left[\left(\frac{L}{L_*}\right)^{p_1} + \left(\frac{L}{L_*}\right)^{p_2}\right]^{-1},
\end{aligned}
\end{eqnarray}
and

\begin{eqnarray}
\label{eqn:flc}
e_2(z)=10^{k_1z+k_2z^2}.
\end{eqnarray}

In equation \ref{local}, $\phi_1$ is the normalization factor of $\rho(z,L)$. The normalization factor of $\Phi(\alpha,z,L)$ is signed as $\phi_0$. Obviously, $\phi_1$ and $\phi_0$ are not independent. They are related by
\begin{eqnarray}
\label{eeeeee}
\int\int\rho(z,L)\frac{dV}{dz}dzdL\equiv \int\int\int\Phi(\alpha,z,L)\frac{dV}{dz}d\alpha dzdL
\end{eqnarray}

Recently, \citet{2012MNRAS.422.2274C} performed a detailed study on spectral properties of radio AGNs. They found that the Gaussian forms can well describe the spectral index distribution for both steep-($\alpha>0.5$) and flat-spectrum ($\alpha<0.5$) sources. It also showed that the standard deviation for flat-spectrum sources is larger than that for steep-spectrum sources (see the Table 2 of their paper). Here we use a Gaussian form of
\begin{eqnarray}
\label{SID}
\frac{dN}{d\alpha}=\exp\left[{-\frac{(\alpha-\mu)^2}{2\sigma^2}}\right]
\end{eqnarray}
to model the intrinsic spectral index distribution $dN/d\alpha$, adopting the same parameters given by \citet{2012MNRAS.422.2274C}. They are [$\mu=0.75, \sigma=0.23$] and [$\mu=0.00, \sigma=0.42$] for steep- and flat-spectrum populations, respectively.

Finally, we summarize the parameters of the input RLFs in table 1.

\begin{figure*}[!htb]
\centering
\includegraphics[width=2.12\columnwidth]{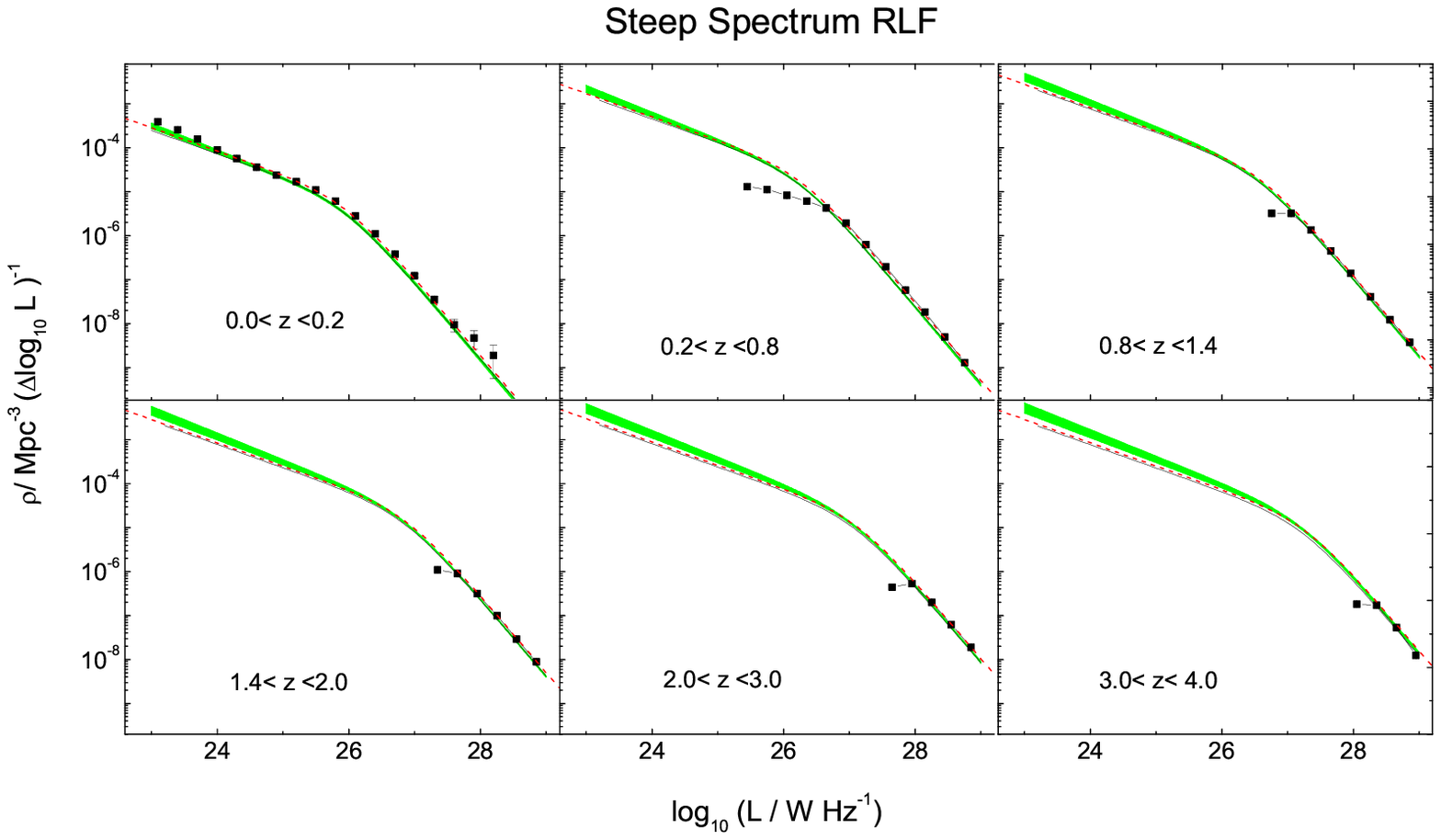}
\includegraphics[width=2.12\columnwidth]{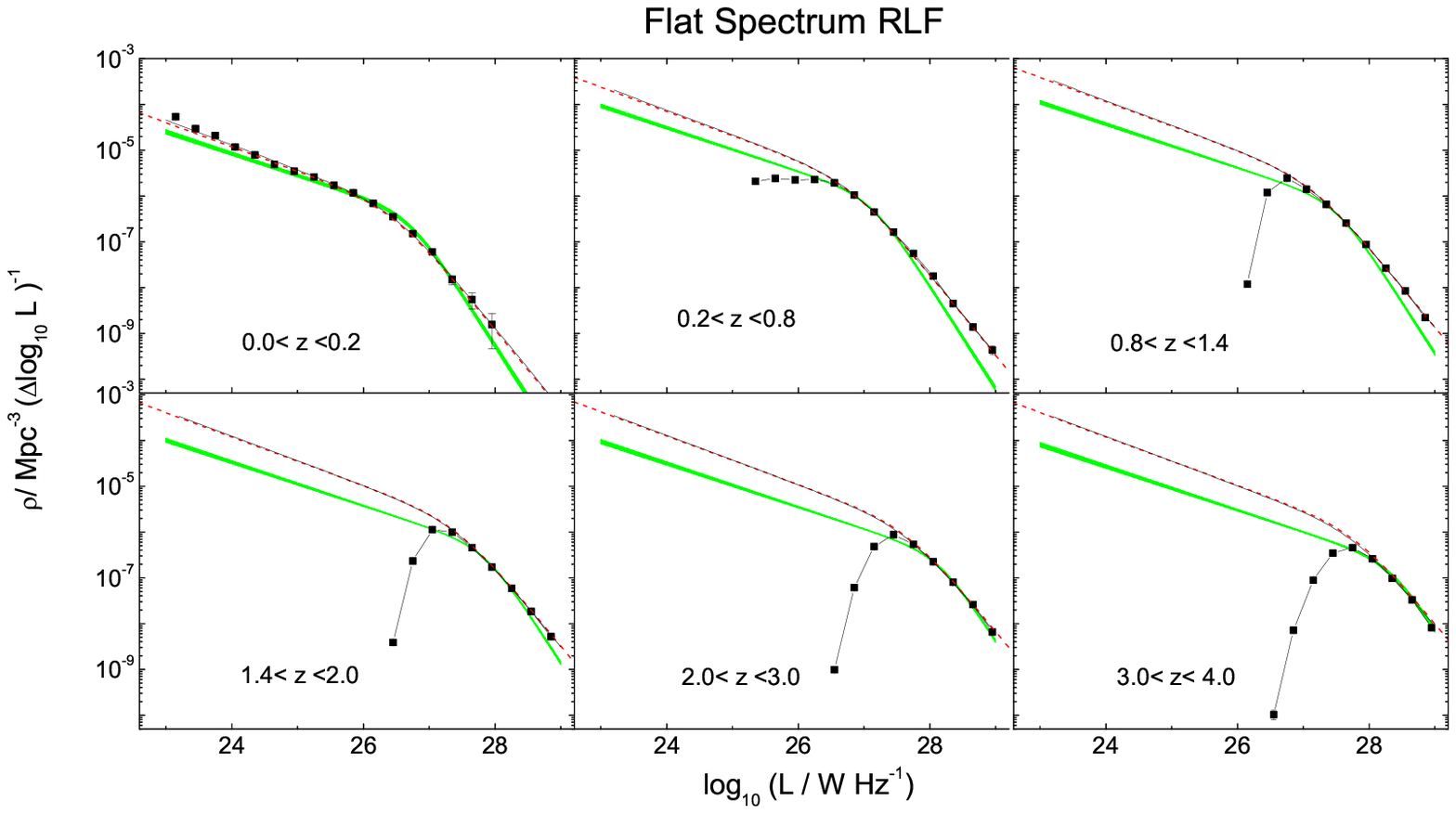}
\caption{RLFs estimated by the three methods described in section 2.4 and 2.5 for steep- and flat-spectrum populations, respectively. The $1/V_a$ estimated RLFs are shown as black solid squares with Poisson errors for six redshift bins. The input RLFs take values of $z = 0.1, 0.35, 0.75, 1.25, 1.75, 2.25$ and are shown as red dashed lines. The green bands show the 2D MLE estimated RLFs with 2$\sigma$ uncertainties. The 3D MLE estimated RLFs are shown as black solid lines, which nearly overlap with the input RLFs.}
\label{RLF-L}
\end{figure*}

\subsection{Monte Carlo simulation}

By inserting equation \ref{eqn:rhoform} and \ref{SID} into \ref{eqpc}, we can obtain random draws of ($\alpha,z,L$) as follows:
\begin{enumerate}
  \item We first draw a random value of $\alpha_i$ from the spectral index distribution $dN/d\alpha$, and then draw a random value of $z_i$ from its marginal distribution $p_z(z)$, which is given by the integral of equation (3) over $\alpha$ and $L$. A random value of $\L_i$ is then drew from the conditional distribution $p(L|\alpha_i,z_i)$. Thus we obtain a simulated source $i$ with $z=z_i$, $\alpha =\alpha_i$, and $L=L_i$.
  \item Derive the flux density $S_i$ of source $i$ by
        \begin{eqnarray}
          \label{eqn:flc}
          L_i=4\pi \times D^2(z)\times S_i /(1+z)^{1-\alpha_i},
        \end{eqnarray}
        where $D(z)$ is the luminosity distance \citep[see][]{1999astro.ph..5116H}.
  \item If $S_i\geqslant S_{lim}$ (the flux-density limit) and $\alpha_i>0.5$, the source $i$ will enter the simulated steep-spectrum sample. For the simulated flat-spectrum sample, the criterion is $S_i\geqslant S_{lim}$ and $\alpha_i<0.5$.
  \item Repeat steps 1-3 $N_{tot}$ times.
\end{enumerate}
In a real simulation process, the value of $N_{tot}$ can be adjusted by changing the simulated solid angle. In addition, we impose the luminosity and redshift respectively to be $0<z<4$ and $23<L<29$ on the simulated sample. Finally we respectively obtain a flux limited ($S_{lim}$=0.2Jy) steep- and flat-spectrum simulated sample containing about two million sources. Figure \ref{simulated_samples} shows two sub-samples containing about 20,000 sources from our simulation.

\begin{figure}[!htb]
\centering
\includegraphics[width=1.22\columnwidth]{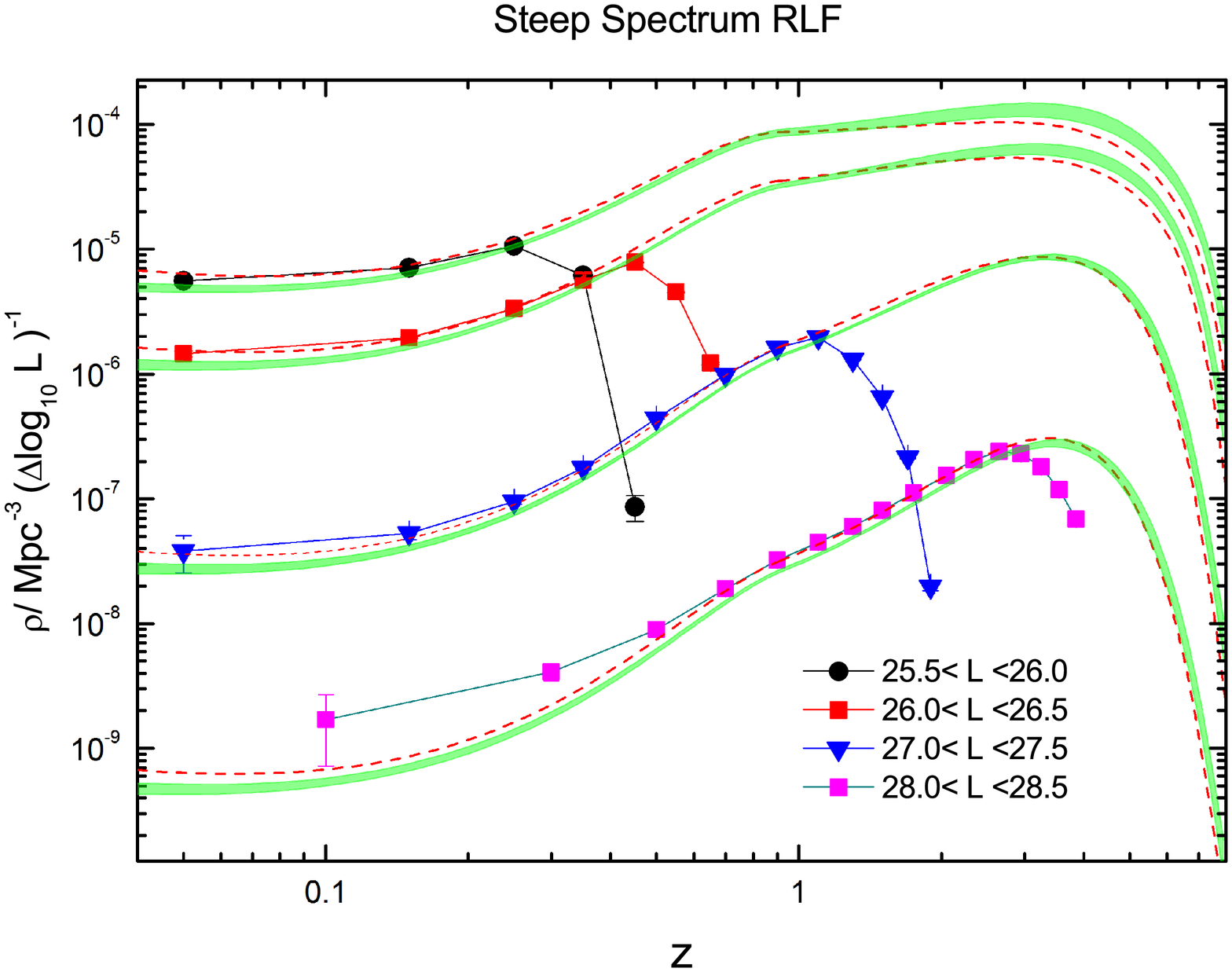}
\includegraphics[width=1.22\columnwidth]{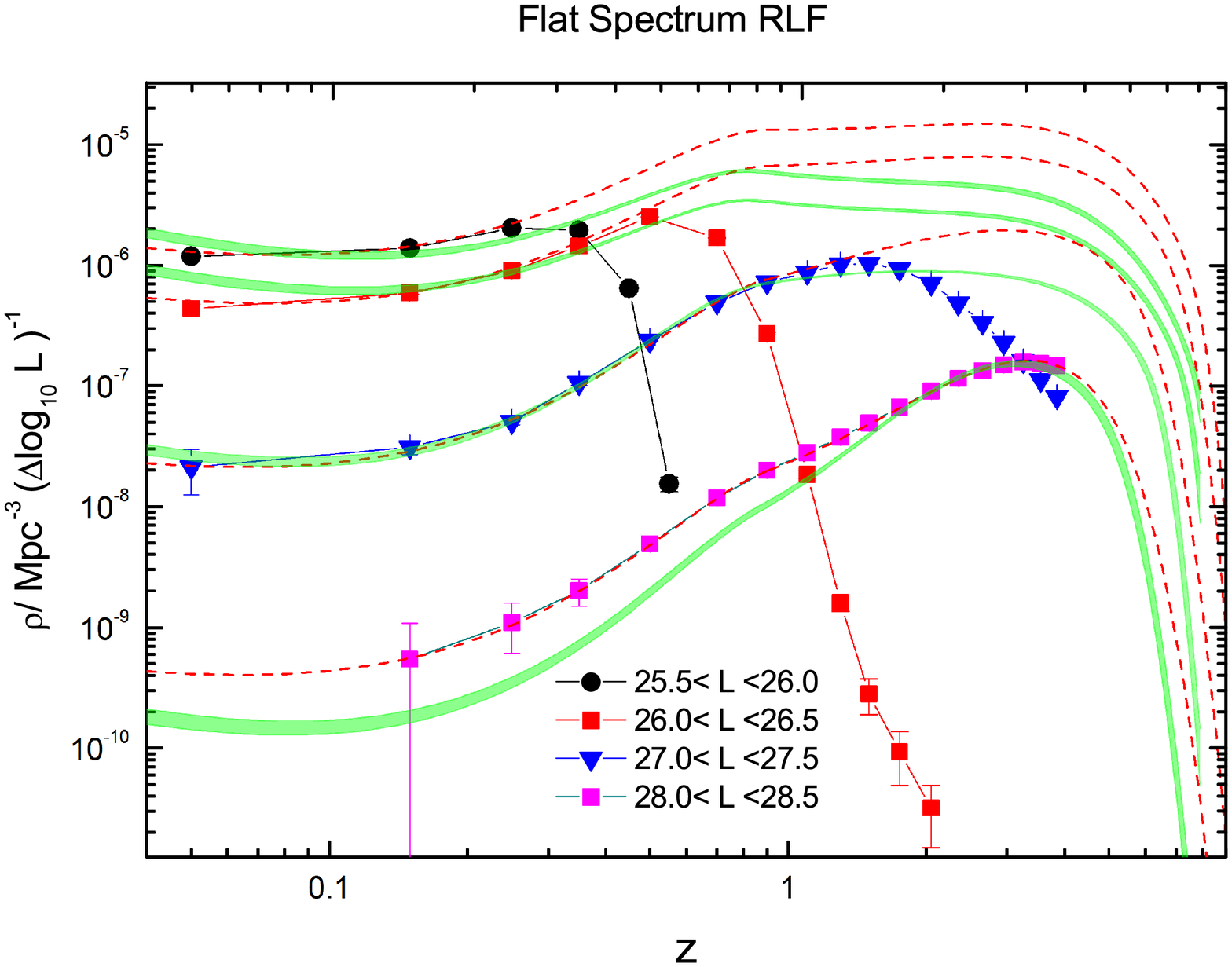}
\caption{Space densities as a function of redshift. The red dashed lines represent the input RLFs at $\log_{10}L$=25.75, 26.25, 27.25, and 28.25 respectively. The green bands show the 2D MLE estimated RLFs with 2$\sigma$ uncertainties. The $1/V_a$ estimated RLFs are shown as colored points with Poisson errors for four power ranges, $\log_{10}L=25.5-26.0,26.0-26.5,27.0-27.5$, and $28.0-28.5$.}
\label{RLF-z}
\end{figure}

\subsection{Estimating the RLF by the traditional methods}

The traditional methods usually treat the RLF as a bivariate density estimation problem, where the spectral index $\alpha$ is not a variable in the function and its effect can only be evaluated indirectly. Now we will investigate how accurate estimation can these traditional methods give for the input RLF defined in section 2.2. The first method tested is the most classical non-parametric $1/V_a$ estimator \citep[see][]{1968ApJ...151..393S,1980ApJ...235..694A}. Although papers after papers \citep[e.g.,][]{2008ApJ...686..148C,2013Ap&SS.345..305Y} have pointed out the bias of this estimator, it is not outdated and continues to be widely used in the literature \citep[see][for latest use of this method]{2016MNRAS.460....2P,2016arXiv160503387Y,2016MNRAS.457..730P}. The key point of the $1/V_a$ method is that it takes into account the contribution of object i to the number density of the bin $\Delta L\Delta z$ as $1/(\Delta L V_{a}^i)$. If $N$ objects appear in the interval $\Delta L \Delta z$ ($L_{1}<L<L_{2}, z_{1}<z<z_{2}$) around the bin center ($L,z$), the LF is estimated as

\begin{eqnarray}
\phi_{1/V_a}(L,z)=\frac{1}{\Delta L}\sum_{i=1}^{N}\frac{1}{V^{i}_{a}}.
\end{eqnarray}

$V_{a}^i$ is the available volume for object $i$, and it is calculated as
\begin{eqnarray}
V_{a}^{i}=\Omega \int_{z_{1}}^{z_{max}^{i}}\frac{dV}{dz}dz.
\end{eqnarray}
If $z(L,\alpha,S)$ is the redshift at which a radio source of luminosity $L$ and spectral index $\alpha$ has a flux density $S$, then
\begin{eqnarray}
z_{max}^{i}=min[z_2,z(L_i,\alpha_i,S_{lim})].
\end{eqnarray}
Obviously, the $1/V_a$ estimator takes into account the spectral index of each individual source. This makes it seem to be trivariate in $\alpha$, $z$ and $L$. However, it only use each individual $\alpha_i$ to calculate $V_{a}^{i}$, but does not give an estimation for the intrinsic spectral index distribution. In this sense, the $1/V_a$ method is still a bivariate estimator.

The second method tested is the maximum likelihood estimate (MLE) method \citep{1983ApJ...269...35M}, specified as two-dimensional (2D) ML method where $\alpha$ is not considered. Given the likelihood function $p(L_{obs},z_{obs}|\theta)$, best estimates for the model parameters are obtained by minimizing $S=-2\ln(p(L_{obs},z_{obs}|\theta))$:
\begin{eqnarray}
\label{likelihood1}
\begin{aligned}
S=-2\sum_{i}^{N_{obs}}&ln[\rho(z_{i},L_{i})]+\\
&2\int\int\rho(z,L)\Omega(z,L)\frac{dV}{dz}dzdL.
\end{aligned}
\end{eqnarray}
For the variable $z$, the lower and  upper limits of the integral are $z_{min}$ and $z_{max}$. For the variable $L$, the upper limit is $L_{max}$, and the lower limit is the flux-density limit line $L=L_{lim}(z)$. However, from Figure 1, we can notice that the flux-density limit line varies with $\alpha$. This situation is particularly notable for the flat-spectrum sample.
We take the average spectral index of the sample as the value of $\alpha$. The $1/V_a$ and 2D MLE estimated results will be given in section 3.

\subsection{Estimating the RLF incorporating the spectral index distribution}

Once the spectral index distribution is incorporated in estimating the RLF, it becomes a trivariate density estimation problem. This is beyond the ability of the existing non-parametric LF estimator. One feasible way is resorting to the parametric method. Here we use the MLE method (specified as three-dimensional MLE, hereafter 3D MLE). Following \citet{2001MNRAS.327..907J} but for the 3-dimensional case, we can write a $\ln$ likelihood function similar with Equation \ref{likelihood1},

\begin{eqnarray}
\label{likelihood2}
\begin{aligned}
S'=-2\sum_{i}^{N_{obs}}&ln[\Phi(\alpha_i,z_{i},L_{i})]+\\
&2\int\int\int\Phi(\alpha,z,L)\Omega(\alpha,z,L)\frac{dV}{dz}d\alpha dz dL.
\end{aligned}
\end{eqnarray}
Considering the limits of the integral in $S'$, we have
\begin{eqnarray}
\label{likelihood3}
\begin{aligned}
S'&=-2\sum_{i}^{N_{obs}}ln[\Phi(\alpha_i,z_{i},L_{i})]+\\
&2\Omega \int^{\alpha_2}_{\alpha_1}d\alpha \int^{z_{2}}_{z_{1}}dz\frac{dV}{dz}\int_{max[L_1,L_{lim}(\alpha,z)]}^{L_2}\Phi(\alpha,z,L)dL,
\end{aligned}
\end{eqnarray}
where $(\alpha_1,\alpha_2)$, $(z_1,z_2)$ and $(L_1^j,L_2^j)$ are limits of spectral index, redshift and luminosity, respectively. $L_{min}(\alpha,z)$ is the luminosity limit surface, and $\Omega$ is the solid angle subtended by the sample. The 3D MLE estimated results will be given in section 3.

\section[]{results}
\subsection{The estimated RLFs}
In Figure \ref{RLF-L}, we show the RLFs estimated by the three methods described in section 2.4 and 2.5 for steep- and flat-spectrum populations, respectively. The $1/V_a$ estimated RLFs are shown as black solid squares with Poisson errors for six redshift bins. The input model RLFs take values of $z = 0.1, 0.57, 1.1, 1.7, 2.4, 3.3$, and are shown as red dashed lines. The green bands show the 2D MLE estimated RLFs with 2$\sigma$ uncertainties, and the 3D MLE estimated RLFs are shown as black solid lines. Generally, the $1/V_a$ and 2D MLE methods give biased estimation, and the bias increases towards high redshift. This is because the $K$-correction increases with redshift. The bias is especially pronounced for flat-spectrum populations whose spectral index distribution is more scattered. The $1/V_a$ estimator is prone to underestimate the faint end RLF and produce an artificial redshift-related flattening or break, giving an impression of luminosity-dependent evolution. Fortunately, the $1/V_a$ estimator can give a good estimation for local RLF (see the redshift bin $0.0<z<0.2$).

The 2D MLE method can give a relatively good estimation to the steep-spectrum RLF, while it only give a poor approximation to the flat-spectrum RLF. We also find that the 2D MLE result closely depends on the choice of flux limit line that is a function of $\alpha$ and redshift. By contrast, the 3D MLE method can give an excellent estimation for both the steep- and flat-spectrum RLFs.

\subsection{RLFs as a function of redshift}
A second presentation of the RLF data is given in Figure \ref{RLF-z}, in which space densities are plotted as a function of redshift for four ranges of intrinsic power. The $1/V_a$ estimated RLFs are shown as colored points with Poisson errors, and the 2D MLE result is shown as green bands with 2$\sigma$ uncertainties. The dashed lines represent the input model RLFs taking values of $\log_{10}L=25.75, 26.25, 27.25, 28.25$. It can be seen clearly that the $1/V_a$ estimator always give an artificial luminosity-dependent decline of space density (known as redshift cut-off in the literature) for bins near the flux cutoff of the sample. Our input model RLF indeed advocates a luminosity-dependent evolution where the low-luminosity radio sources experience weak evolution to $\thicksim$ 1, while the high-luminosity radio sources undergo strong evolution to $z>3$. However, the $1/V_a$ estimator obviously exaggerate the significance/degree of this luminosity-dependent evolution, producing an impression that the space density of the powerful sources peaks at higher redshift than that of their weaker counterparts.

\begin{figure}[!htb]
\centering
\includegraphics[width=1.22\columnwidth]{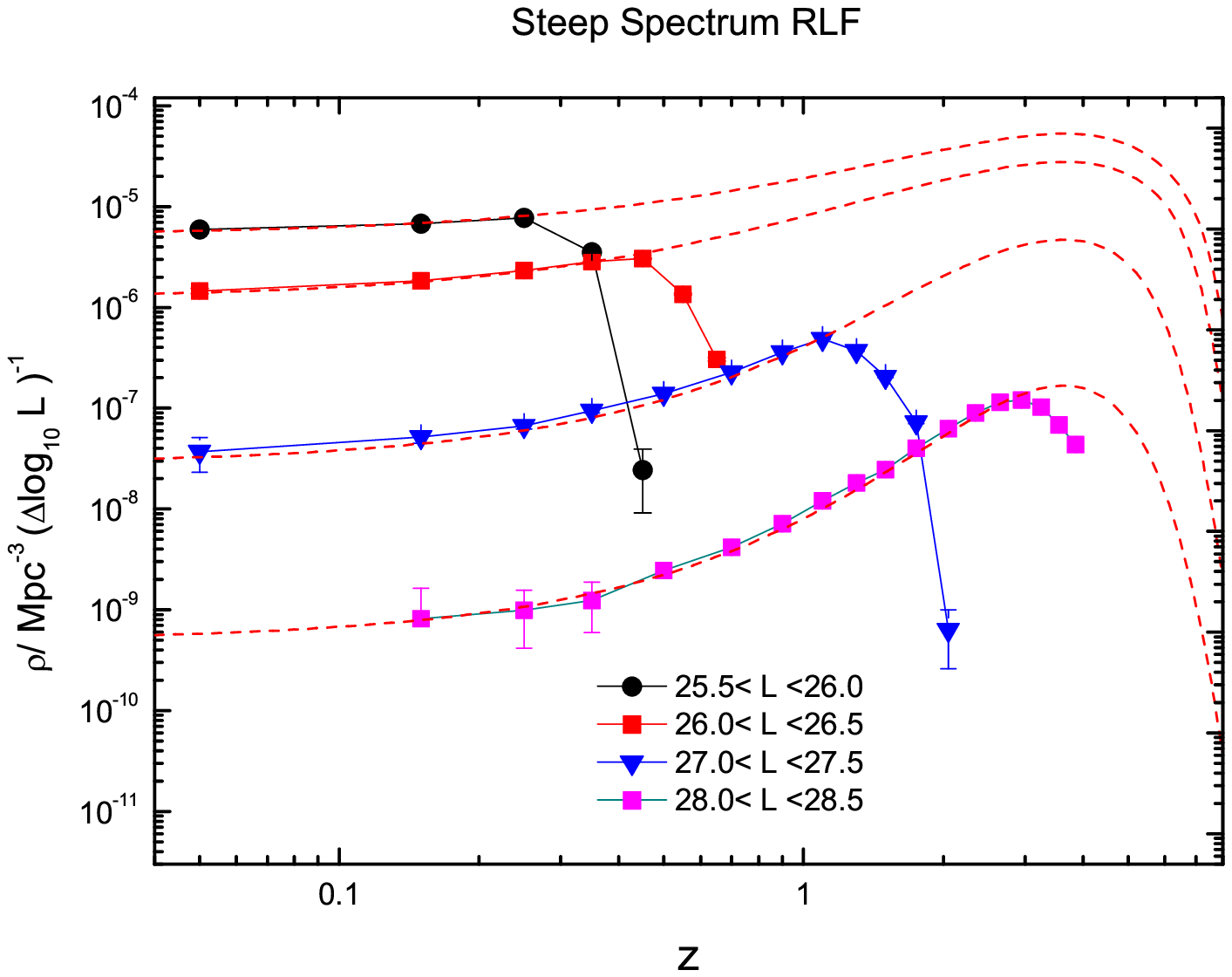}
\includegraphics[width=1.22\columnwidth]{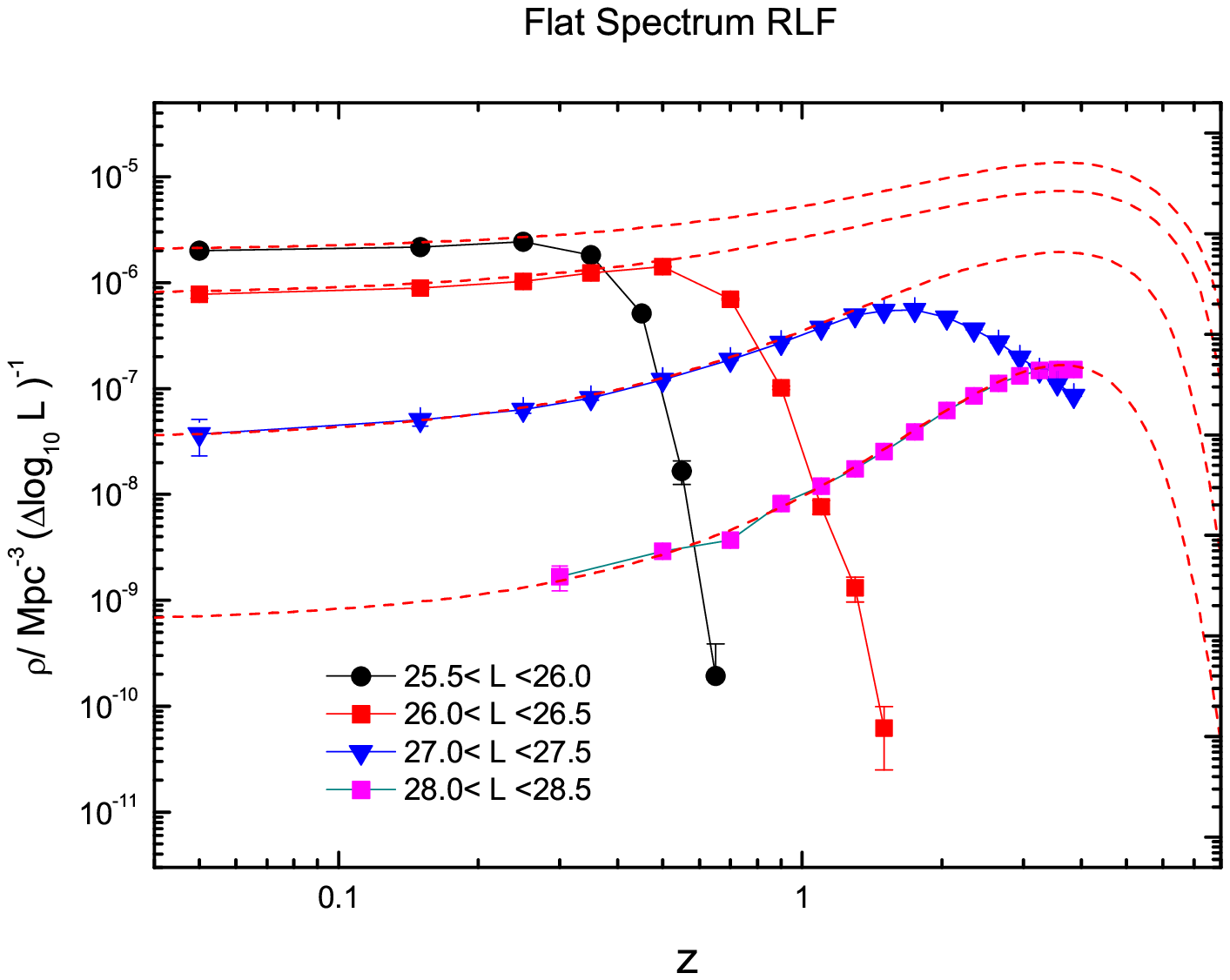}
\caption{Space densities as a function of redshift for another simulated sample. The red dashed lines represent the input RLFs at $\log_{10}L$=25.75, 26.25, 27.25, and 28.25 respectively. The $1/V_a$ estimated RLFs are shown as colored points with Poisson errors for four power ranges, $\log_{10}L=25.5-26.0,26.0-26.5,27.0-27.5$, and $28.0-28.5$.}
\label{PLE}
\end{figure}

\section[]{Discussion}
\subsection{The reason of bias}
It is found that the bivariate RLF estimators can cause significant bias. The primary reason is that in the analysis process of bivariate RLF estimators, the effect of spectral index distribution is not sufficiently considered. In a real radio survey, one often make great efforts to construct a so-called ``flux-limited complete" sample. In fact, one also potentially attach a spectral index selection criterion to the final sample. Take the steep-spectrum sample for example, only the sources with spectral index ($\alpha>0.5$) are selected into the final sample. Thus the selection function of radio sample is actually a two-dimensional surface defined as $L=L_{lim}(\alpha,z)$ in the $L-z-\alpha$ space. The robustness of a LF estimator depends on how accurately it can deal with the selection function (or referred as boundary condition) of the truncated data. Due to the intrinsic scatter of spectral index as well as the fact that $K$-correction is a function of both spectral index and redshift, the truncation boundary on the $L-z$ plane of sample is complicated. The traditional bivariate RLF estimators have difficulty in dealing with this boundary condition properly. The 2D MLE simply treat the boundary condition as a one-dimensional curve but not a two-dimensional surface. For the $1/V_a$ estimator, it does not stick to a single survey-limit line in the $L-z$ plane. It calculate the accessible volume $V_a$ using the spectral properties of each source individually \citep[named as ``source-by-source analysis" by][]{2005A&A...434..133W}. However, this process can neither guarantee that the boundary condition is properly treated. Thus we suggest that a feasible way to get a robust RLF estimation is incorporating the spectral index distribution into the analysis process.

From Figure 1, it is clear that sources near the flux limit become sparse for both the steep- and flat-spectrum samples. This is not a real decline of density, but just a result of selection effect. A bivariate RLF estimator may fail to properly account for this selection effect, and regard the sparseness as a real decline in density. The situation is especially severe for the flat-spectrum radio sources just because their spectral index distribution is more scattered. This may partly explain why a sharp redshift cut-off is easier to observed in the flat-spectrum RLFs \citep[][]{1990MNRAS.247...19D,1996Natur.384..439S,2005A&A...434..133W}, while in the steep-spectrum RLFs it still has much controversy \citep[e.g.,][]{2001MNRAS.327..907J,2007MNRAS.375.1349C,2011MNRAS.416.1900R,2016ApJ...820...65Y}. Therefore, for the flat-spectrum population, it particularly need to incorporate the spectral index distribution into the analysis process to get a robust RLF estimation.

\subsection{Nonparametric- vs. parametric estimator}

Generally, methods of estimating LFs can be classified into nonparametric- and parametric estimators. The nonparametric method constructs the LFs directly from the data and makes no assumptions about the form of the LFs. However, the traditional nonparametric methods are not good at treating the truncation boundary, especially when the spectral index distribution is incorporated and the estimation becomes a trivariate density estimation problem. What's more, multiple samples from different surveys with different flux limits are usually combined together to determining the LFs \citep[e.g.,][]{2011MNRAS.416.1900R}. This case further complicates the process because there exist multiple truncation boundaries. It is easier for the parametric estimators to consider the truncation boundaries, even if for a trivariate density estimation problem. However, the parametric estimators need to assume a particular analytical form for the LFs. This is not an easy task, especially in the absence of any guidance from astrophysical theory. Notably, in recent years some more rigorous approaches, such as a powerful semi-parametric approach \citep{2007ApJ...661..703S} and a Bayesian approach \citep{2008ApJ...682..874K} have been proposed.
Nevertheless, although powerful, these methods are bivariate estimators. They need to be developed to be competent to solve the trivariate density estimation problem. This will be the subject of our future work.

\subsection{More simulation}

In order to rule out the possibility that our input RLF is too special to make the bivariate RLF estimator produce bias, we need to perform more simulation based on other input RLFs. For example, Figure \ref{PLE} shows the result of another simulation in which the input RLF is modeled as a pure luminosity evolution. Again, the $1/V_a$ estimator give an artificial luminosity-dependent decline of space density, exaggerating the significance of luminosity-dependent evolution. In recent years, researches have suggested that the position of RLF peak is luminosity dependent, being interpreted as a sign of cosmic downsizing. This work is not against the robustness of luminosity-dependent evolution in AGN RLFs. But for some previous results, which are based on bivariate RLF estimators and did not sufficiently consider the effect of spectral index distribution, we do worry there is a risk that the significance/degree of luminosity-dependent evolution should be more or less magnified.

\section[]{Conclusions}

The main results of this work are as follows:

\begin{enumerate}
  \item Based on a Monte Carlo method, we find that the traditional bivariate RLF estimators can cause bias in varying degree. The bias is especially pronounced for the flat-spectrum radio sources whose spectral index distribution is more scattered. The bias is caused because the $K$-corrections complicate the truncation boundary on the $L-z$ plane of the sample, but the traditional bivariate RLF estimators have difficulty in dealing with this boundary condition properly. We suggest that the spectral index distribution should be incorporated into the RLF analysis process to obtain a robust estimation.

  \item The classical nonparametric $1/V_a$ estimator can produce an artificial luminosity-dependent decline of space density for bins near the flux cutoff of the sample. As a result of this distortion to the true RLF, the significance/degree of luminosity-dependent evolution is magnified, making a redshift cut-off easier to find.

\end{enumerate}

\acknowledgments

We are grateful to the referee for useful comments that improved this paper. We acknowledge the financial support from the National Natural Science Foundation of China 11133006, 11163006, 11173054, 11573060, the Policy Research Program of Chinese Academy of Sciences (KJCX2-YW-T24), the CAS ``Light of West China" Program, the Strategic Priority Research Program, and the Emergence of Cosmological Structures of the Chinese Academy of Sciences (grant No. XDB09000000). J.Mao is supported by the Hundred-Talent Program of Chinese Academy of Sciences, the Key Research Program of Chinese Academy of Sciences (grant No. KJZD-EW-M06), and the Introducing Overseas Talent Plan of Yunnan Province. The authors gratefully acknowledge the computing time granted by the Yunnan Observatories, and provided on the facilities at the Yunnan Observatories Supercomputing Platform.

\listofchanges


\begin{thebibliography}{}

\bibitem[Ajello et al.(2012)]{2012ApJ...751..108A} Ajello, M., Shaw, M.~S., Romani, R.~W., et al.\ 2012, \apj, 751, 108


\bibitem[Avni \& Bahcall(1980)]{1980ApJ...235..694A} Avni, Y., \& Bahcall, J.~N.\ 1980, \apj, 235, 694


\bibitem[Boyle et al.(2000)]{2000MNRAS.317.1014B} Boyle, B.~J., Shanks, T., Croom, S.~M., et al.\ 2000, \mnras, 317, 1014

\bibitem[Cara \& Lister(2008)]{2008ApJ...686..148C} Cara, M., \& Lister, M.~L.\ 2008, \apj, 686, 148-154

\bibitem[Chhetri et al.(2012)]{2012MNRAS.422.2274C} Chhetri, R., Ekers, R.~D., Mahony, E.~K., et al.\ 2012, \mnras, 422, 2274

\bibitem[Cruz et al.(2007)]{2007MNRAS.375.1349C} Cruz, M.~J., Jarvis, M.~J., Rawlings, S., \& Blundell, K.~M.\ 2007, \mnras, 375, 1349

\bibitem[Dunlop \& Peacock(1990)]{1990MNRAS.247...19D} Dunlop, J.~S., \& Peacock, J.~A.\ 1990, \mnras, 247, 19


\bibitem[Hogg(1999)]{1999astro.ph..5116H} Hogg, D.~W.\ 1999, arXiv:astro-ph/9905116


\bibitem[Ilbert et al.(2004)]{2004MNRAS.351..541I} Ilbert, O., Tresse, L., Arnouts, S., et al.\ 2004, \mnras, 351, 541


\bibitem[Jarvis \& Rawlings(2000)]{2000MNRAS.319..121J} Jarvis, M.~J., \& Rawlings, S.\ 2000, \mnras, 319, 121


\bibitem[Jarvis et al.(2001)]{2001MNRAS.327..907J} Jarvis, M.~J., Rawlings, S., Willott, C.~J., et al.\ 2001, \mnras, 327, 907


\bibitem[Kelly et al.(2008)]{2008ApJ...682..874K} Kelly, B.~C., Fan, X., \& Vestergaard, M.\ 2008, \apj, 682, 874-895

\bibitem[Lake et al.(2016)]{2016arXiv160407493L} Lake, S.~E., Wright, E.~L., Tsai, C.-W., \& Lam, A.\ 2016, arXiv:1604.07493

\bibitem[Marshall et al.(1983)]{1983ApJ...269...35M} Marshall, H.~L., Tananbaum, H., Avni, Y., \& Zamorani, G.\ 1983, \apj, 269, 35


\bibitem[Miyaji et al.(2000)]{2000A&A...353...25M} Miyaji, T., Hasinger, G., \& Schmidt, M.\ 2000, \aap, 353, 25


\bibitem[Peacock(1985)]{1985MNRAS.217..601P} Peacock, J.~A.\ 1985, \mnras, 217, 601

\bibitem[Pracy et al.(2016)]{2016MNRAS.460....2P} Pracy, M.~B., Ching, J.~H.~Y., Sadler, E.~M., et al.\ 2016, \mnras, 460, 2

\bibitem[Prescott et al.(2016)]{2016MNRAS.457..730P} Prescott, M., Mauch, T., Jarvis, M.~J., et al.\ 2016, \mnras, 457, 730

\bibitem[Rigby et al.(2011)]{2011MNRAS.416.1900R} Rigby, E.~E., Best, P.~N., Brookes, M.~H., et al.\ 2011, \mnras, 416, 1900


\bibitem[Schafer(2007)]{2007ApJ...661..703S} Schafer, C.~M.\ 2007, \apj, 661, 703


\bibitem[Schmidt(1968)]{1968ApJ...151..393S} Schmidt, M.\ 1968, \apj, 151, 393


\bibitem[Shaver et al.(1996)]{1996Natur.384..439S} Shaver, P.~A., Wall, J.~V., Kellermann, K.~I., Jackson, C.~A., \& Hawkins, M.~R.~S.\ 1996, \nat, 384, 439


\bibitem[Wall et al.(2005)]{2005A&A...434..133W} Wall, J.~V., Jackson, C.~A., Shaver, P.~A., Hook, I.~M., \& Kellermann, K.~I.\ 2005, \aap, 434, 133


\bibitem[Willott et al.(2001)]{2001MNRAS.322..536W} Willott, C.~J., Rawlings, S., Blundell, K.~M., Lacy, M., \& Eales, S.~A.\ 2001, \mnras, 322, 536


\bibitem[Yuan \& Wang(2013)]{2013Ap&SS.345..305Y} Yuan, Z., \& Wang, J.\ 2013, \apss, 345, 305


\bibitem[Yuan \& Wang(2012)]{2012ApJ...744...84Y} Yuan, Z., \& Wang, J.\ 2012, \apj, 744, 84

\bibitem[Yuan et al.(2016)]{2016ApJ...820...65Y} Yuan, Z., Wang, J., Zhou, M., \& Mao, J.\ 2016, \apj, 820, 65

\bibitem[Yuan et al.(2016)]{2016arXiv160503387Y} Yuan, Z.~S., Han, J.~L., \& Wen, Z.~L.\ 2016, arXiv:1605.03387

\bibitem[Zeng et al.(2013)]{2013MNRAS.431..997Z} Zeng, H., Yan, D., \& Zhang, L.\ 2013, \mnras, 431, 997


\end{thebibliography}
\end{document}